\documentclass[showpacs,prd,prd,nofootinbib,floatfix,amsmath,amssymb]{revtex4}
\usepackage{graphicx}
\begin{document}

%\tightenlines

\title{Production of doubly charged vector bilepton pairs at $\gamma \gamma$ colliders}
\author{C. G. Honorato}
\email[E-mail:]{est073@fcfm.buap.mx}
\author{J. J. Toscano}
\email[E-mail:]{jtoscano@fcfm.buap.mx} \affiliation{Facultad de
Ciencias F\'\i sico Matem\'aticas, Benem\'erita Universidad
Aut\'onoma de Puebla, Apartado Postal 1152, Puebla, Pue., M\'
exico}

\date{\today}

\begin{abstract}
The production of pairs of doubly charged vector bileptons is
studied at future $\gamma \gamma$ colliders. The unpolarized
cross--section for the $\gamma \gamma \to Y^{--}Y^{++}$ subprocess
is analytically calculated and convoluted to predict the number of
events in the complete $e^+e^-\to \gamma \gamma \to Y^{--}Y^{++}$
process. The gauge or non--gauge character of the vector bilepton
$Y^{\pm \pm}$ is discussed. It is found that as a consequence of
its spectacular signature, as it decays dominantly into two
identical charged leptons, and also due to its charge contents,
which significantly enhance the cross--section, the detection of
this class of particles with mass in the sub--TeV region can be at
the reach of these colliders. The model--independent nature of our
results is stressed.
\end{abstract}

\pacs{14.70.Pw, 12.20.-m}

\maketitle

\section{Introduction}
There are many reasons to believe that new physics beyond the
Fermi scale must exist, but it remains unclear just how or where
it will observed. However, it is quite likely that signals of new
physics would be more evident in those processes which are
forbidden or strongly suppressed in the standard model (SM). One
possible source of new physics effects could be associated with
processes that violate global conservation laws, such as, for
instance, lepton number, since there is no reason to believe in
its exactness. One peculiarity of the standard model (SM) is that
none of its bosons carry global quantum numbers. As a consequence,
lepton number is conserved both separately and globally. However,
many SM extensions predict physical processes that can violate
such conservation laws. For instance, the separate lepton number
conservation is violated by bileptons, which are scalar or vector
particles that carry two units of lepton number. This class of
particles can arise in many well motivated SM
extensions\cite{Review}. For instance, scalar bileptons are
present in theories with enlarged Higgs sectors, as left--right
symmetric models\cite{LRM} in which appear doubly charged scalars
coupling to pairs of identical leptons. As far as non--gauge
massive vectors are concerned, they are present in
composite\cite{Composite} and technicolor\cite{Technicolor}
theories. On the other hand, massive gauge bileptons can appear
when the SM gauge group is embedded into a larger gauge group,
such as occurs in the $SU(15)$ unified model\cite{GUT} or in the
so called $331$ models\cite{331}, which are based in the
$SU_C(3)\times SU_L(3)\times U_X(1)$ group.

In this work, we study the production of pairs of doubly charged
vector bileptons $Y^{\pm \pm}$ in high energy $\gamma \gamma$
collisions with the photons originating from Compton laser
backscattering. The $\gamma \gamma \to Y^{\pm \pm }Y^{\mp \mp}$
reaction is interesting for several reasons. First, there is the
distinctive signature of this class of bileptons, as they
predominantly decay into two identical charged leptons, {\it i.e.}
$Y^{\pm \pm}\to 2l^\pm$, which constitutes an almost SM
background--free signal. In fact, a possible source of SM
background--free signal is the process $\gamma \gamma \to
Z^*W^{-*}W^{+*}\to 4l^{\pm}\bar{\nu}_l\nu_l$, which evidently is
very suppressed. Secondly, this process can be studied in a
model--independent manner, since it is entirely governed by the
electromagnetic $U_e(1)$ symmetry. In this respect, we will
introduce the most general dimension--four $U_e(1)$--invariant
Lagrangian. As already mentioned, one interesting peculiarity of a
vector bilepton is the fact that it could be a gauge field (GF) or
a non--gauge vector field (from now on we will refer to it simply
as a Proca Field (PF)), which appears reflected in the
high--energy behavior of the cross--section. The GF or PF nature
of the vector bilepton will be incorporated in the
$U_e(1)$--invariant Lagrangian and its phenomenological
implications discussed. Finally, a third reason to study this
process is the fact that its cross section is larger than that
associated with a singly charged particle by a facto of
$Q^4_Y=2^4=16$. On the other hand, the prospects for the detection
of bileptons at future colliders have been extensively studied in
the literature by several authors. Since bileptons couple mainly
to leptons, most works have been focused on colliders involving at
least one lepton beam. The potential of next linear colliders
operating in the same charge mode to produce doubly charged
bileptons as a $s$--channel resonance has received special
attention\cite{C,P1,G}. In particular, the $e^-e^-\to Y^{--}\to
\mu^-\mu^-$ process is specially suited for search of doubly
charged GF, since their couplings to leptons are large and
prescribed\cite{P1}. These colliders can also be used to study
doubly charged scalar resonances\cite{G}, though it should be
noted that in this case their couplings to leptons are free
parameters and are small, as they arise from the Yukawa sector.
Most studies on GF bilepton production have been done in the
context of the minimal $331$ model, which predicts the existence
of a pair of singly and doubly charged gauge bosons, $Y^\pm_\mu$
and $Y^{\pm \pm}_\mu$\cite{BZ}. Diverse production mechanisms have
been analyzed at hadron\cite{Hadron}, $e^-p$\cite{LeptonHadron},
$e^+e^-$\cite{Lepton}, $e^-\gamma$\cite{LeptonPhoton}, and
muon\cite{MUON} colliders. Though single vector bilepton
production has been studied at $\gamma \gamma$
colliders\cite{LONDON}, to our knowledge, the issue of pairs
production at this type of colliders only has been tackled
marginally\cite{Review}. In this paper, we will present a
comprehensive study of the $\gamma \gamma \to Y^{--}Y^{++}$
process, which includes the derivation of explicit analytical
expressions for the cross section of the $\gamma \gamma \to
Y^{--}Y^{++}$ subprocess. In addition to dealing with both the GF
and PF cases, our results will be applicable to a wide range of
models, as they were derived by invoking only the general
principles of renormalization theory, Lorentz invariance, and
$U_e(1)$--gauge invariance.

The paper has been organized as follows. In Sec. \ref{Lag}, the
most general Lorentz and $U_e(1)$--invariant Lagrangian, which
includes only interactions up to dimension--four, is presented. In
Sec. \ref{Cal}, the cross section for the complete $e^+e^-\to
\gamma \gamma \to Y^{\pm \pm}Y^{\mp \mp}$ reaction is derived. In
Secs. \ref{Dis} and \ref{Con}, we discuss our results and present
the conclusions, respectively.

\section{The $U_e(1)$--invariant Lagrangian and Feynman rules}
\label{Lag}In this section, we discuss the general structure of
the couplings between a doubly charged vector bilepton and the
electromagnetic gauge field. The most general Lagrangian including
interactions up to dimension--four, which respects the Lorentz and
electromagnetic symmetries, can be written as
\begin{eqnarray}
{\cal L}&=&-\frac{1}{2}(D_\mu Y^{++ }_\nu-D_\nu Y^{++}_\mu)^\dag
(D^\mu Y^{++\nu}-D^\nu Y^{++\mu})+m^2_YY^{--}_\mu
Y^{++\mu}\nonumber \\
&&-ieQ_Y\kappa F_{\mu \nu}Y^{--\mu}Y^{++\nu}-\frac{1}{4}F_{\mu
\nu}F^{\mu \nu},
\end{eqnarray}
where $D_\mu=\partial_\mu -ieQ_YA_\mu$ and $F_{\mu
\nu}=\partial_\mu A_\nu-\partial_\nu A_\mu$ are the covariant
derivative and the field strength tensor associated with the
$U_e(1)$ gauge group, respectively. In addition, $Q_Y=2$ is the
electric charge of the bilepton in units of the positron charge
and $\kappa$ is a dimensionless parameter, which is related with
the CP--even on--shell electromagnetic properties of the $Y^{\pm
\pm}$ vector field, namely, the magnetic dipole moment and the
electric quadrupole moment\cite{TT}. The good high--energy
behavior of the theory and therefore its renormalizability depends
critically on the precise value of this parameter. The $\kappa=0$
case corresponds to a PF, which can be present in effective field
frameworks, such as technicolor thoeries\cite{TCL}. On the other
hand, GF bileptons are identified with $\kappa=1$. In this case,
the $F_{\mu \nu}Y^{--\mu}Y^{++\nu}$ term arises just from a
Yang--Mills Lagrangian, such as the one associated with the $331$
models\cite{BZ}. The possibility of a parameter $\kappa=1+\Delta
\kappa$, with $\Delta \kappa \ll 1$, cannot be ruled out in the
case of a GF, as the $\Delta \kappa$ correction can be associated
with anomalous contributions arising from radiative corrections
within the context of a wider theory. This type of deviations are
well--known from the SM $W$ boson, as they have been widely
studied within the context of electroweak effective
Lagrangians\cite{EW}. Although all these possibilities are in
principle of physical interest, it is important to stress that the
amplitude associated with the $\gamma \gamma \to Y^{--}Y^{++}$
scattering has a good high--energy behavior only for
$\kappa=1$\cite{GHB}. Bellow we will retain the $\kappa$ parameter
until we proceed with the analysis of the cross section for the
complete $e^+e^-\to \gamma \gamma \to Y^{--}Y^{++}$ process.

\begin{figure}
\centering
\includegraphics[width=2.5in]{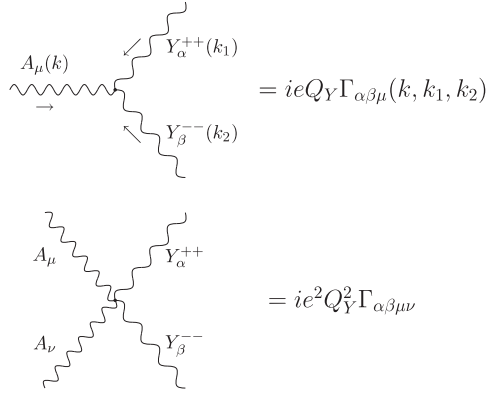}
\caption{\label{FIG1}Feynman rules for the $Y^{--}Y^{++}\gamma$
and $Y^{--}Y^{++}\gamma \gamma$ vertices.}
\end{figure}

We now turn to present the Feynman rules for the
$Y^{--}Y^{++}\gamma$ and $Y^{--}Y^{++}\gamma \gamma$ vertices. Our
notation and conventions are shown in Fig.\ref{FIG1}, where the
$\Gamma_{\alpha \beta \mu}(k,k_1,k_2)$ and $\Gamma_{\alpha \beta
\mu \nu}$ Lorentz tensors are given by
\begin{equation}
\Gamma_{\alpha \beta \mu}(k,k_1,k_2)=(k_1-k_2)_\mu g_{\alpha
\beta}+(k_2-\kappa k)_\alpha g_{\beta \mu}-(k_1-\kappa k)_\beta
g_{\alpha \mu},
\end{equation}
\begin{equation}
\Gamma_{\alpha \beta \mu \nu}=-2g_{\alpha \beta}g_{\mu
\nu}+g_{\alpha \mu}g_{\beta \nu}+g_{\alpha \nu}g_{\beta \mu}.
\end{equation}

\section{The cross section}
\label{Cal}
\begin{figure}
\centering
\includegraphics[width=3in]{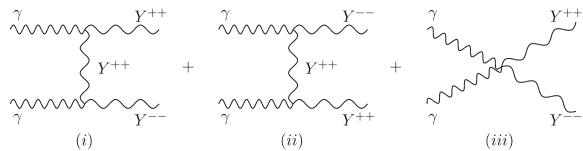}
\caption{\label{FIG2}Feynman diagrams contributing to the $\gamma
\gamma \to Y^{--}Y^{++}$ reaction.}
\end{figure}

In this section, we calculate the cross section
associated with the complete $e^+ e^-\to \gamma \gamma \to
Y^{--}Y^{++}$ process. We first discuss the properties of the
$\gamma \gamma \to Y^{--}Y^{++}$ subprocess. We will use the
following notation
\begin{equation}
A_\mu (k_1)+A_\nu (k_2)\to Y^{++}_\alpha (k_3)+Y^{--}_\beta (k_4),
\end{equation}
where $k_1+k_2=k_3+k_4$. With this convention, the corresponding
Mandelstam variables are given by $\hat{s}=(k_1+k_2)^2$,
$\hat{t}=(k_1-k_3)^2$, and $\hat{u}=(k_2-k_3)^2$, which satisfy
the relation $\hat{s}+\hat{t}+\hat{u}=2m^2_Y$. In Fig.\ref{FIG2},
we show the Feynman diagrams contributing to the invariant
amplitude, which can be written as
\begin{equation}
{\cal M}=-4\pi i\alpha Q^2_Y{\cal M}_{\mu \nu \alpha
\beta}\epsilon^\mu (k_1,\lambda_1)\epsilon^\nu
(k_2,\lambda_2)\epsilon^{\alpha *} (k_3,\lambda_3)\epsilon^{\beta
*} (k_4,\lambda_4),
\end{equation}
where $\epsilon^\mu (k_1,\lambda_1)$, $\epsilon^\nu
(k_2,\lambda_2)$ etc., represent the polarization four--vectors
associated with the external particles. As a consequence of
electromagnetic gauge--invariance, the ${\cal M}_{\mu \nu \alpha
\beta}$ tensorial amplitude obeys the following Ward identities:
\begin{equation}
k^\mu_1 {\cal M}_{\mu \nu \alpha \beta}=k^\nu_2 {\cal M}_{\mu \nu
\alpha \beta}=0.
\end{equation}
It turns out to be that in the most general case, {\it i.e.} for
an arbitrary $\kappa$ parameter, the amplitude is characterized by
six independent tensor structures which satisfy the gauge
conditions. So, the tensorial amplitude can be written as follows:
\begin{equation}
{\cal M}_{\mu \nu \alpha \beta}=\sum^6_{i=1}f_i N^i_{\mu \nu
\alpha \beta},
\end{equation}
where the $ N^i_{\mu \nu \alpha \beta}$ are the tensor structures,
which are given by
\begin{eqnarray}
 N^1_{\mu \nu \alpha \beta}&=&(1+3\kappa)\Big(-\frac{\hat{s}}{2}g_{\mu
 \alpha}g_{\nu \beta}+k_{1\alpha}k_{2\mu}g_{\nu
 \beta}-k_{1\alpha}k_{2\beta}g_{\mu \nu}+k_{2\beta}k_{1\nu}g_{\mu
 \alpha}\Big), \\
N^2_{\mu \nu \alpha
\beta}&=&(1+3\kappa)\Big(-\frac{\hat{s}}{2}g_{\nu
 \alpha}g_{\mu \beta}+k_{2\alpha}k_{1\nu}g_{\mu
 \beta}-k_{2\alpha}k_{1\beta}g_{\mu \nu}+k_{1\beta}k_{2\mu}g_{\nu
 \alpha}\Big), \\
 N^3_{\mu \nu \alpha \beta}&=&4g_{\alpha \beta}\Bigg(\frac{g_{\mu
 \nu}}{2f_1f_2}+\hat{s}k_{3\mu}k_{3\nu}-\frac{k_{2\mu}k_{3\nu}}{f_1}-\frac{k_{3\mu}k_{1\nu}}{f_2}\Bigg),
 \\
 N^4_{\mu \nu \alpha
 \beta}&=&2(1+\kappa)\Bigg[\Big(k_{2\alpha}g_{\beta
 \nu}-k_{2\beta}g_{\alpha
 \nu}\Big)\Bigg(\hat{s}k_{3\mu}-\frac{k_{2\mu}}{f_1}\Bigg)+\Big(k_{1\alpha}g_{\mu
 \beta}-k_{1\beta}g_{\mu
 \alpha}\Big)\Bigg(\hat{s}k_{3\nu}-\frac{k_{1\nu}}{f_2}\Bigg)\Bigg],\\
 N^5_{\mu \nu \alpha
 \beta}&=&(1-\kappa)\Big(2f_1k_{1\alpha}k_{3\mu}-g_{\mu
 \alpha}\Big)\Big[g_{\nu \beta}+2f_1k_{2\beta}(k_3-k_1)_\nu \Big],
 \\
 N^6_{\mu \nu \alpha
 \beta}&=&(1-\kappa)\Big(2f_2k_{2\alpha}k_{3\nu}-g_{\nu
 \alpha}\Big)\Big[g_{\mu \beta}+2f_2k_{1\beta}(k_3-k_2)_\mu \Big].
\end{eqnarray}
In the above expressions, the $f_i$ Lorentz scalars are given by
\begin{eqnarray}
f_1&=&\frac{1}{m^2_Y-\hat{t}}, \ \ \ f_2=\frac{1}{m^2_Y-\hat{u}}, \\
f_3&=&f_4=f_1f_2, \\
f_5&=&\frac{1}{4m^2_Yf_1}, \ \ \ f_6=\frac{1}{4m^2_Yf_2}.
\end{eqnarray}
Notice that the number of tensor structures reduces to four for a
gauge bilepton without the presence of anomalous static
electromagnetic properties ($\kappa=1$).

The unpolarized cross section for the $\gamma \gamma \to
Y^{--}Y^{++}$ subprocess is given by
\begin{equation}
\hat{\sigma}(\gamma \gamma \to Y^{--}Y^{++})=\frac{1}{16\pi^2
\hat{s}^2}\int^{\hat{t}_2}_{\hat{t}_1}d\hat{t} |\bar{{\cal M}}|^2,
\end{equation}
where the integration limits are given by
\begin{equation}
\hat{t}_2(\hat{t}_1)=-\frac{\hat{s}}{4}\Bigg(1\pm
\sqrt{1-\frac{4m^2_Y}{\hat{s}}}\Bigg).
\end{equation}
After solving the integral, one obtains
\begin{equation}
\hat{\sigma}(\gamma \gamma \to Y^{--}Y^{++})=\frac{\pi
\alpha^2Q^4_Y}{\hat{s}}F(x,\kappa),
\end{equation}
where $x=4m^2_Y/\hat{s}$ and
\begin{eqnarray}
F(x,\kappa)&=&\Big(-3(\kappa-481)x^5+(223\kappa-1952)x^4+3(1733\kappa+1179)x^3\nonumber
\\ &&+(10175-9407\kappa)x^2
+4(8533\kappa-2389)x-2400(\kappa-1)\Big){\cal A}_1(x)\nonumber
\\
&&+\Bigg(\frac{24x^3-4(\kappa+1)x^2+(1-\kappa)x-22(\kappa-1)}{8x}\Bigg){\cal
A}_2(x).
\end{eqnarray}
The ${\cal A}_i$ functions appearing in the above expression are
given by:
\begin{eqnarray}
{\cal A}_1(x)&=&\frac{\sqrt{1-x}}{48x^2(x^3-2x^2-7x+24)}, \\
{\cal A}_2(x)&=&Log\Bigg[\frac{4(1+\sqrt{1-x})-x(x-1)}{
4(1-\sqrt{1-x})-x(x-1)}\Bigg],
\end{eqnarray}
Notice that, as it could be expected, $F(1,\kappa)=0$.

With the photons originating from Compton laser backscattering,
the unpolarized cross--section for the complete $e^+e^-\to \gamma
\gamma \to Y^{--}Y^{++}$ process is given by
\begin{equation}
\sigma(s)=\int^{y_{max}}_{2m^2_Y/\sqrt{s}}dz\frac{d{\cal
L}_{\gamma \gamma}}{dz}\hat{\sigma}(\gamma \gamma \to
Y^{--}Y^{++}),
\end{equation}
where $\hat{s}=z^2s$, with $\sqrt{s}(\sqrt{\hat{s}})$ the center
of mass energies of the $e^+e^-(\gamma \gamma)$ collisions, and
$\frac{d{\cal L}_{\gamma \gamma}}{dz}$ is the photon luminosity,
defined as
\begin{equation}
\label{l1} \frac{d{\cal L}_{\gamma
\gamma}}{dz}=2z\int^{y_{max}}_{z^2/y_{max}}\frac{dy}{y}f_{\gamma/e}(y)f_{\gamma/e}(z^2/y),
\end{equation}
where the energy spectrum of the back scattered photon is given
by\cite{FUN}
\begin{equation}
\label{l2}
f_{\gamma/e}(y)=\frac{1}{D(\chi)}\Big[1-y+\frac{1}{1-y}-\frac{4y}{\chi
(1-y)}+\frac{4y^2}{\chi^2(1-y)^2}\Big],
\end{equation}
with
\begin{equation}
\label{l3}
D(\chi)=\Big(1-\frac{4}{\chi}-\frac{8}{\chi^2}\Big)Log(1+\chi)+\frac{1}{2}+\frac{8}{\chi}-\frac{1}{2(1+\chi)^2}.
\end{equation}
In this expression, $\chi=(4E_0\omega_0)/m^2_e$, where $m_e$ and
$E_0$ are the mass and energy of the electron, respectively;
$\omega_0$ is the laser--photon energy, and $y$ represents the
fraction of the energy of the incident electron carried by the
backscattered photon. The optimum values for the $y_{max}$ and
$\chi$ parameters are $y_{max}\approx 0.83$ and
$\chi=2(1+\sqrt{2})$. Notice that the $x$ and $z$ variables are
related through $x=4m^2_Y/sz^2$.

It has been customary to treat numerically the $\frac{d{\cal
L}_{\gamma \gamma}}{dz}$ photon luminosity, but we prefer to solve
it analytically. After solving this integral given through
Eqs.(\ref{l1},\ref{l2},\ref{l3}), one obtains
\begin{equation}
\frac{d{\cal L}_{\gamma
\gamma}}{dz}=\frac{4z}{D(\chi)^2(z^2-1)^4\chi^4}\Bigg[g_0(z)+g_1(z)Log(y_{max})+g_2(z)Log(z)+
g_3(z)Log\Big(\frac{z^2-y_{max}}{y_{max}-1}\Big)\Bigg],
\end{equation}
where
\begin{eqnarray}
g_0(z)&=&-16y_{max}(y^2_{max}-z^2)\Big(z^4+\chi
z^2(z^2-1)\Big)\nonumber \\
&&+4\chi^2y_{max}(y^2_{max}-z^2)(z^6-3z^4+4z^2-2)\nonumber \\
&&-2\chi^4(y_{max}-1)(z^2-1)^2\Big(z^4-y_{max}z^2(y_{max}+1)+\chi^3\Big),\\
g_1(z)&=&2\chi^2(z^2-1)^3\Big(\chi (2z^2+\chi +4)+4\Big),\\
g_2(z)&=&-8\chi^3z^4(z^2-1)^2+16z^4(z^2+1)+32\chi
z^4(z^2-1)\nonumber \\
&&+8\chi^2z^2(z^4-3z^2+2)+\chi^4(z^4-3z^3+2)^2,\\
g_3(z)&=&-16z^4(z^2+1)-32(z^2-1)-\chi^4(z^2-1)^2(z^4-2z^2+2)\nonumber
\\
&&+4\chi^3(z^2-1)^2(z^4-z^2+2)-8\chi^3(2z^6-6z^4+5z^2-1).
\end{eqnarray}

The total cross section $\sigma(s)$ cannot be expressed in closed
form and will be numerically evaluated in the next section.

\section{Results and discussion}
\label{Dis}The next linear colliders (NLC) $e^+e^-$ in the TeV
region, including its derivations $\gamma e$ and $\gamma \gamma$,
will open up new opportunities in particle physics\cite{GP}.
Although these colliders are intended to operate initially at a
center of mass energy of a few hundreds of GeVs with a luminosity
of the order of $10^{33}$ $cm^{-2}s^{-1}$, it is contemplated to
increase the energy up to about $2000$ GeV in subsequent stages.
In particular, the $\gamma \gamma$ collisions offer the
opportunity of accessing to physical information not available
from $e^+e^-$ colliders. Although the same type of particles can
be produced in these colliders, it is important to stress that the
reactions are different and thus complementary information can be
obtained. One important feature of the $\gamma \gamma$ colliders
is that the cross sections are much larger than in $e^+e^-$
collisions. For instance, the cross section of $W$ pair production
in the $\gamma \gamma$ collision is almost two orders of magnitude
higher than the $W$ pair production in $e^+e^-$ collisions.
Another important difference between $e^+e^-$ and $\gamma \gamma$
collisions is that, while the cross sections in the former
decreases for increasing energies, those associated with the
latter are almost constant. One disadvantage of a $\gamma \gamma$
collider is that its energy reaches approximately $80\%$ of the
parent $e^+e^-$ collider. Thus, $e^+e^-$ colliders operating in
the region of a few TeVs will be necessary in order to investigate
the production of new particles with masses in the sub--TeV
region.

Having briefly discussed the main features of $\gamma \gamma$
colliders, we proceed to discuss our results. To begin with, it is
convenient to comment on the bounds on the doubly charged vector
bilepton mass. Currently, the most stringent bound arises from the
conversion of muonium ($\mu^- e^+$) to antimuonium ($\mu^+ e^-$),
which leads to the limit $m_Y>1417 g_l$ GeV\cite{MAM}, where $g_l$
is a model dependent constant coupling characterizing the
$\bar{l}^clY^{++}$ vertex, which is expected to be smaller than
$1$. However, it has been argued that this bound can be evaded in
a more general context since it relies on very restrictive
assumptions\cite{PPF}. Another strong limit, $m_Y>1250g_l$ GeV,
arises from fermion pair production and lepton flavor violating
decays\cite{TULLY}. Recently, a more realistic constraint,
$m_Y>577g_l$, was derived from Bhabha scattering with CERN LEP
data\cite{LEP}. We would like to stress that all the above bounds
are model dependent, and so the existence of lighter doubly
charged vector bileptons is still allowed. We will present results
for a bilepton with mass in the range $2m_W<m_Y<10m_W$, with $m_W$
the SM $W$ gauge boson mass.

\begin{figure}
\centering
\includegraphics[width=4in]{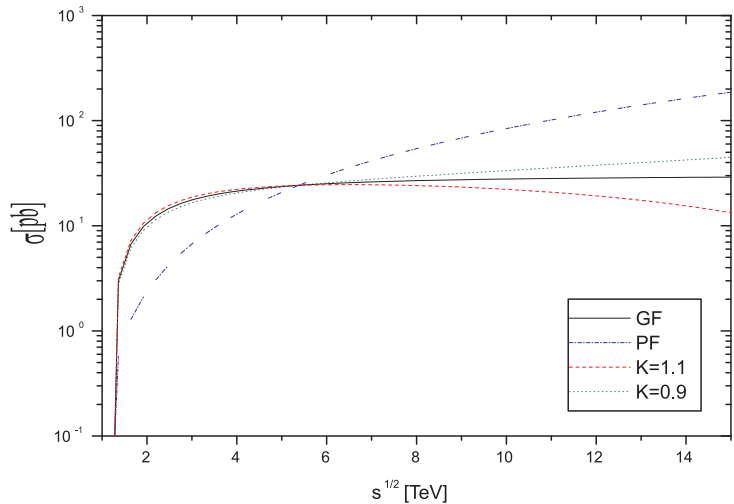}
\caption{\label{FIG3}The cross section for the $e^+e^-\to \gamma
\gamma \to Y^{--}Y^{++}$ reaction as a function of the center of
mass energy for $m_Y=450$ GeV. Contributions of PF ($\kappa=0$)
and GF ($\kappa=1$), including anomalous effects ($\kappa=0.9$ and
$\kappa=1.1$), are shown.}
\end{figure}

Before analyzing our results for some $\sqrt{s}$ energies of
$e^+e^-$ colliders, it is interesting to investigate the
high--energy behavior of the cross section. In Fig.\ref{FIG3}, we
present the $\sigma (e^+e^-\to \gamma \gamma \to Y^{--}Y^{++})$
cross section as a function of $\sqrt{s}$ for both PF ($\kappa=0$)
and GF bileptons ($\kappa=1$), including anomalous effects that
represent deviations of $10\%$ with respect to the renormalizable
value of $\kappa$. As already mentioned, the former case defines
vector electrodynamics, which, as it is well--known, is not a
predictive theory. The latter case, corresponds to a $F_{\mu
\nu}Y^{--\mu}Y^{++\nu}$ coupling induced by a Yang--Mills strength
tensor $F^a_{\mu \nu}$. From this figure, it can be appreciated
that the cross section associated with GF without anomalous
contributions has a good high--energy behavior, as it tends to a
constant value for increasing energies. As already mentioned, this
essentially constant behavior of the cross section for the case of
a GF is a peculiarity of $\gamma \gamma$ colliders. In contrast,
in the case of PF, the corresponding cross section is
ill--behavior at high energies. As far as the anomalous effects
are concerned, it can be appreciated that, for the values
$m_Y=450$ GeV and $\Delta \kappa =\pm 0.1$, the cross section is
sensitive only for energies higher than about one order of
magnitude than the bielepton mass. It can also be observed that,
for increasing energies, the cross section decreases for $\kappa
>1$ and increases for $\kappa <1$. These results suggest that an
unmoderated growth of the cross section is expected in the range
$0\leq \kappa <1$, reaching its highest ill--behavior with the
energy for a PF ($\kappa=0$). These results agree with the
well--known fact that only Yang--Mills theories ($\kappa=1$) lead
to well--behaved cross sections of binary processes involving
massive vector bosons\cite{GHB}.

\begin{figure}
\centering
\includegraphics[width=6.5in]{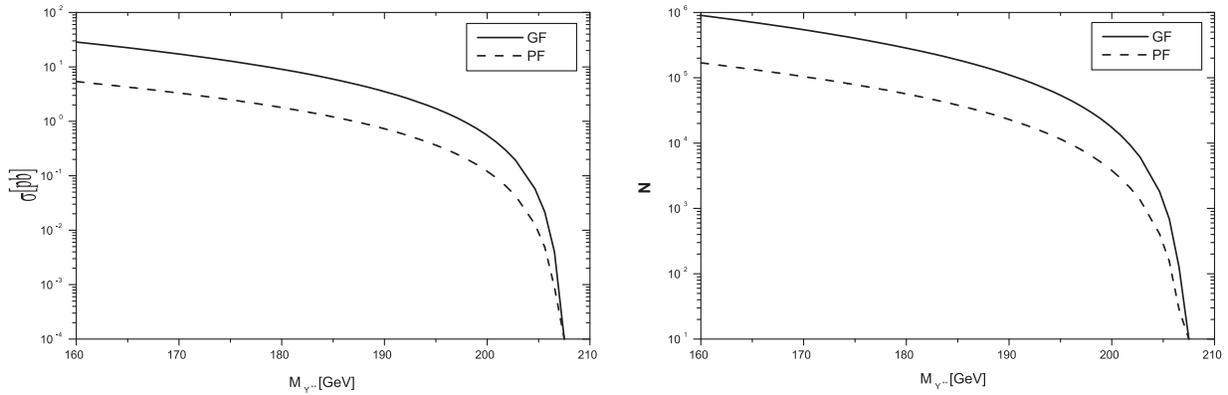}
\caption{\label{FIG4} The cross section and the number of events
($N$) as a function of $m_Y$ for $\sqrt{s}=0.5$ TeV.}
\end{figure}

\begin{figure}
\centering
\includegraphics[width=6.5in]{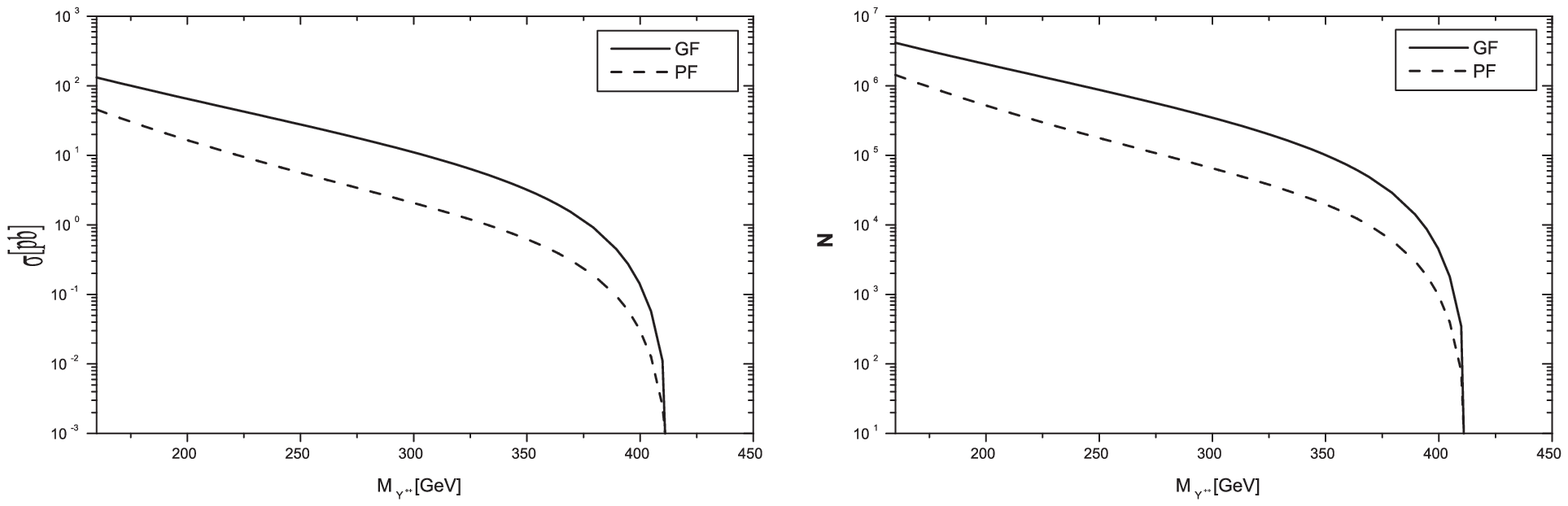}
\caption{\label{FIG5} The same as Fig.\ref{FIG4}, but now for
$\sqrt{s}=1$ TeV.}
\end{figure}

\begin{figure}
\centering
\includegraphics[width=6.5in]{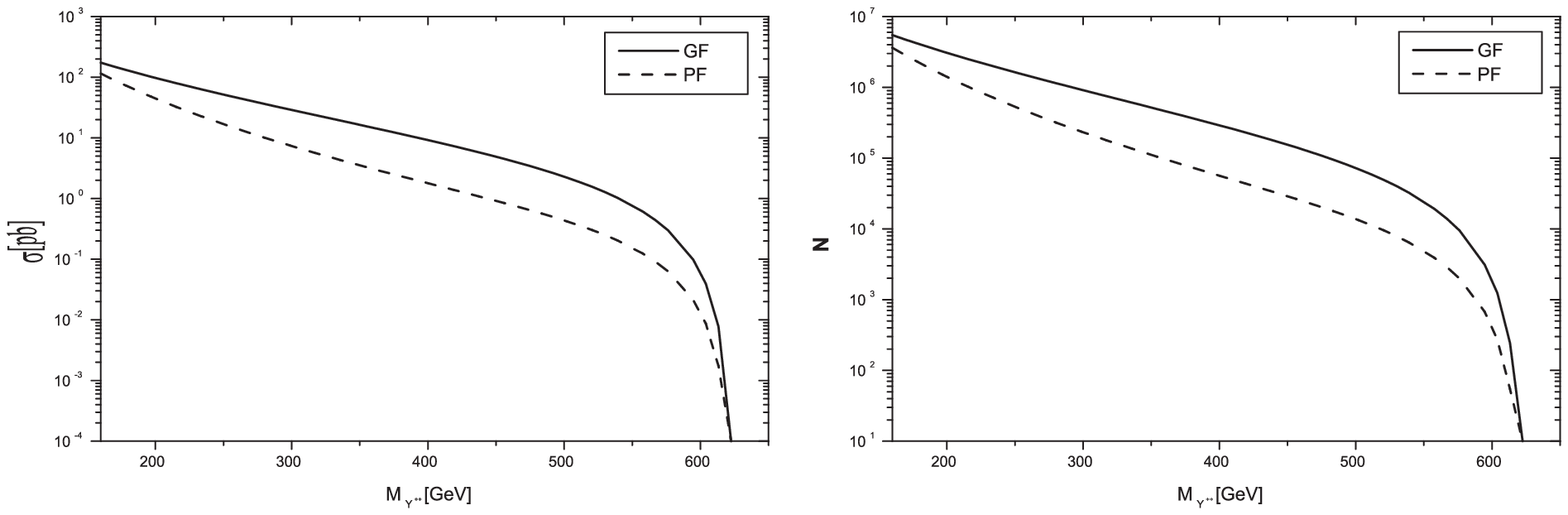}
\caption{\label{FIG6} The same as Fig.\ref{FIG4}, but now for
$\sqrt{s}=1.5$ TeV.}
\end{figure}

\begin{figure}
\centering
\includegraphics[width=6.5in]{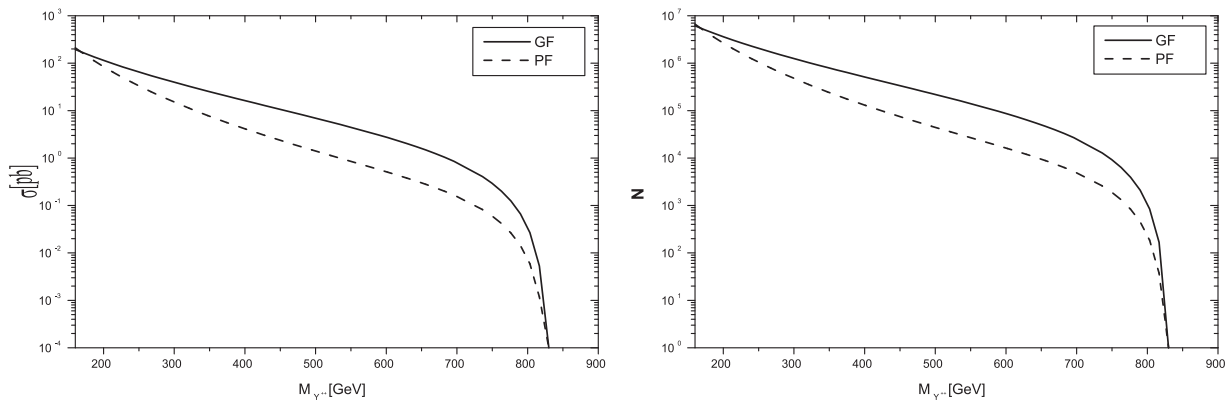}
\caption{\label{FIG7} The same as Fig.\ref{FIG4}, but now for
$\sqrt{s}=2$ TeV.}
\end{figure}

We now turn to discuss our results for some scenarios. We will
analyze only the PF and GF cases, as they represent two extreme
situations. We will present results for the cross section $\sigma$
and the number of events $N$ using an integrated luminosity of
$10$ $fb^{-1}$ and energies of $\sqrt{s}=0.5, 1, 1.5$, and $2$
TeV. In Fig.\ref{FIG4}, the number of events and the cross section
are presented as functions of $m_Y$ and for $\sqrt{s}=0.5$ TeV. We
can see that for a relatively light bilepton, with mass in the
range $2m_W<m_Y<200$ GeV, the cross section for a GF ranges
between almost $29$ pb and $0.55$ pb, and it is approximately $5$
times larger than that of the PF. With the luminosity considered,
the number of events of GF bileptons ranges between $9 \times
10^5$ and $1.7 \times 10^4$. As it occurs with the cross sections,
the number of events of GF and the number of events of PF differ
in the same proportion. Indeed, the
$\sigma_{GF}/\sigma_{NGF}=N_{GF}/N_{NGF}=R<1$ ratios are true for
any energy, as it can be observed from Figs.\ref{FIG4}-\ref{FIG7}.
It is also evident from these figures that $R$ decreases for
increasing energies, as $\sigma_{NGF}$ increases whereas at the
same time $\sigma_{GF}$ remains essentially constant. Indeed, in
the range $0.5<\sqrt{s}<2$ TeV, $R$ is always smaller than $1$, as
it can be appreciated from Fig.\ref{FIG3}. In the following, we
will concentrate on the number of events and only for the case of
GF bileptons. Thus, from Fig.\ref{FIG5}, we can see that $N$
ranges between  $4\times 10^6$ and $10^5$ for $\sqrt{s}=1$ TeV and
$2m_W<m_Y<350$ GeV. On the other hand, from Fig.\ref{FIG6}, we can
appreciate that, for $\sqrt{s}=1.5$ TeV, $N$ goes from $5.5 \times
10^6$ for $m_Y=2m_W$ to $7.2\times 10^4$ for $m_Y=500$ GeV.
Finally, it is found that, for $\sqrt{s}=2$ TeV, $N$ goes from
$6.2\times 10^6$ for $m_Y=2m_W$ to $10^3$ for $m_Y=800$ GeV.

It is interesting to compare our results with the case of $W$ pair
production. Using the same integrated luminosity for the
convoluted cross section, the number of events is approximately
$10^6$, $1.5\times 10^6$, $1.7\times 10^6$, and $1.8\times 10^6$
for energies of $0.5$, $1$, $1.5$, and $2$ TeV, respectively. The
number of events estimated in the literature for a nonconvoluted
cross section, at energies above $200$ GeV and for an integrated
luminosity of $100$ $fb^{-1}$, is about $8\times 10^6$\cite{GP}.

\section{Conclusions}
\label{Con}A $\gamma \gamma$ collider is the ideal place to study
the production of pairs of new charged particles and their
properties, as the cross sections are large and model independent
at the tree level. In this work, we have studied the production of
pairs of doubly charged vector bileptons, whose decay modes
$Y^{\pm \pm}\to 2l^\pm$ will provide a spectacular signature,
since it is almost free of SM background. The unpolarized cross
section for the $\gamma \gamma \to Y^{- - }Y^{+ +}$ subprocess was
analytically derived and then convoluted to predict the cross
section and the number of events of the complete $e^+e^-\to \gamma
\gamma \to Y^{-- }Y^{++}$ process. Our results are valid for an
arbitrarily charged vector boson that includes anomalous couplings
characterized by the dimension--four operator $ie\kappa F_{\mu
\nu}Y^{--\mu}Y^{++\nu}$, with $\kappa$ an real arbitrary
parameter. The renormalizable case ($\kappa=1$), which corresponds
to a gauge bilepton, as well as the possibility of a matter vector
($\kappa=0)$, which corresponds to vector electrodynamics, were
analyzed. The possibility of a gauge bilepton with anomalous
interactions was studied too. Two specific values were considered,
namely, $\kappa=0.9$ and $\kappa=1.1$. It was shown that, as it is
well--known, the cross section has a high--energy well--behavior
only for the case $\kappa=1$, and it is essentially constant for
increasing energies. On the other hand, for the case of a
nonrenormalizable Lagrangian, it was found that for increasing
energies, the cross section increases for values of $\kappa$
ranging from $0$ to $1$, but decreases for $\kappa>1$. The worst
of all the behaviors is found for $\kappa=0$. The situation is
reversed when the cross section is analyzed as a function of the
bielepton mass $m_Y$. In this case, for a given energy, the
largest and smallest number of events is found for $\kappa=1$ and
$\kappa=0$, respectively. Our conclusion is that this class of
exotic particles can be detected still if they are as heavy as
about $10m_W$, as the number of events is relatively large, of
about $10^3$, and their signature is spectacular.

\acknowledgments{We appreciate useful discussions with G.
Tavares--Velasco. Support from Conacyt and VIEP--BUAP (M\' exico)
is acknowledged.}

\end{document}